\documentclass[preprint,aps,showpacs]{revtex4}
\usepackage{graphicx}
\usepackage{amsmath}

\begin{document}

\title{Generation of Low Frequency Plasma Waves After Wave-Breaking}

\author{Prabal Singh Verma\footnote{prabal-singh.verma@univ-amu.fr}}
\affiliation{CNRS/Universit\'e d'Aix-Marseille
Centre saint J\'er\^ome, case 232, F-13397 Marseille cedex 20, France
}

\date{\today}
\begin{abstract}
Spatio-temporal evolution of a non-relativistic electrostatic waves in a cold plasma has been studied in the 
wave-breaking regime using a 1D particle-in-cell simulation.  
It is found that plasma gets heated after the wave-breaking but 
a fraction of initial energy always remains with the remnant wave in the form of BGK mode in warm plasma. 
An interesting finding of this work is that 
the frequency of the resultant BGK wave is found be below electron plasma 
frequency which decreases with increasing initial amplitude. 
Moreover, the acceleration mechanism after the wave-breaking is also found to be different from 
the previous work. In order to explain the results observed in the numerical experiments, a simplified theoretical model 
is constructed which exhibits a good agreement with the simulation.  
These investigations have direct relevance in wake-field acceleration experiments.      

\end{abstract}


\maketitle
\section{Introduction}

Breaking of electrostatic waves \cite{dawson1959nonlinear, wang1992wave, davidson1968nonlinear,davidson1972methods, rowlands2008exact, 
brodin2017simple} 
in plasma has a wide range of application from wake-field acceleration  
\cite{modena1995electron,faure2004laser,caldwell2009proton,nature2014} to plasma heating in nuclear fusion experiments \cite{kodama2001fast}. 
The phenomenon of wave-breaking was originally introduced by Dawson \cite{dawson1959nonlinear}  
using a sheet model in the cold plasma approximation where thermal corrections have been ignored. 
Considering ions which are much more massive than electrons fixed in space, Dawson has shown that as long as 
amplitude of the perturbation is small, oscillations stay 
linear and electron sheets oscillate coherently at electron plasma frequency. 
However, at larger amplitudes frequency of the oscillations remains 
the same but a peak in the electron density and corresponding steepening in the electric field profile starts to emerge as a nonlinear 
feature of the electron-plasma oscillations. At the critical amplitude (wave-breaking amplitude) electron density peak theoretically 
goes to infinity and coherent electrostatic energy starts to convert into random kinetic energy. It was predicted that 
all the electrostatic energy would get converted into random kinetic energy once the coherent oscillations are broken 
\cite{dawson1959nonlinear,wang1992wave}.  
In contrast it was shown in the previous work that a fraction of energy always remains with the wave as 
a superposition of two BGK-modes \cite{verma2012residual}. These modes form dynamically over several plasma periods after the breaking 
of standing cold plasma oscillations initiated by a sinusoidal density perturbations for $\delta n_e /n_0 > 0.5$.  

The present work aims at investigating the behaviour of nonlinear plasma waves in the wave breaking regime 
because this configuration is more interesting from the application point of view.  
These waves can be excited in the wake of the laser pulse  
when it propagates into the plasma, provided frequency of the laser pulse matches the electron-plasma frequency \cite{gibbon2004short}.  
Note here that when the 
group velocity of the laser pulse is close to speed of light,  
the wake waves are described by the time stationary solution of the relativistic electron fluid equations 
\cite{akhiezer1956theory,infeld1989relativistic,verma2012breaking,bera2015fluid}. However, if the the group velocity of the 
laser pulse is much smaller than speed of light, wake waves in principle are described by the time-stationary 
solution of nonrelativistic electron fluid equations \cite{davidson1968nonlinear,davidson1972methods,albritton1975relation}.  
An exact time-stationary solution for such nonlinear waves has been first suggested by Davidson and Schram 
\cite{davidson1968nonlinear,davidson1972methods}. These waves are also known as cold plasma BGK modes 
which can be obtained from the Lagrange solution too, for an unique choice of initial conditions 
\cite{albritton1975relation}.  
When the maximum velocity amplitude of these waves becomes greater than or equal to the phase velocity $\omega_{pe}/k$,  
wave-breaking occurs due to the trajectory crossing between neighbouring oscillating electrons forming the wave  
and destroys the coherent motion. 
Therefore, from the wake-field acceleration point of view it is interesting to investigate 
what happens to these modes after they are broken. Do 
they also form coherent structures after the wave breaking as shown in the previous work for the case of 
standing oscillations ?  If it is so, what kind of distribution functions are formed ? 
And most importantly what is the acceleration mechanism that allows the wave to capture and accelerate slower particles 
in order to loose coherent energy ? Note that in the case of standing plasma oscillation 
different `$k$' modes have different phase velocities which were playing the main role in the acceleration process 
after the wave-breaking \cite{verma2012residual}.  However, that is not the case for traveling waves because here  
all the `$k$' modes propagate with same phase velocity $\sim \omega_{pe}/k$.
Therefore, the aim of present work is to 
answer aforementioned questions  
in an effort to gain more insight into the physics of nonlinear plasma waves in the wave-breaking 
regime during their spatio-temporal evolution.  

We use one dimensional particle-in-cell (1D PIC) simulations \cite{birdsall2004plasma} to study the space-time 
evolution of cold plasma BGK 
wave \cite{davidson1968nonlinear,davidson1972methods,albritton1975relation} in the wave-breaking regime and find that the plasma gets heated after the wave-breaking but 
all the coherent energy does not vanish completely  
as a fraction of it always remains with the wave in the form of a single BGK mode in warm plasma 
\cite{bernstein1957exact,hutchinson2017electron}. 
Although, the result is consistent with the previous finding \cite{verma2012residual}, interestingly, 
the frequency of the final BGK wave is found to be below electron plasma frequency which  
decreases with increasing initial amplitude. Besides, the acceleration mechanism after the wave-breaking 
is also found to be very different from previous work \cite{verma2012residual}. In order to provide an understanding of this 
behaviour we construct a simplified theoretical model which shows a good agreement with numerical results.     

The paper is organized as follows. Section II contains results from the simulation. In section III, interpretation 
of the numerical results is provided and section IV describes the theoretical modeling. 
Section V contains the summary and discussion of all the results presented in this paper. 

\section{Results from the simulation}
\begin{figure}
\begin{minipage}[b]{0.45\linewidth}
\centering
\includegraphics[width=\textwidth]{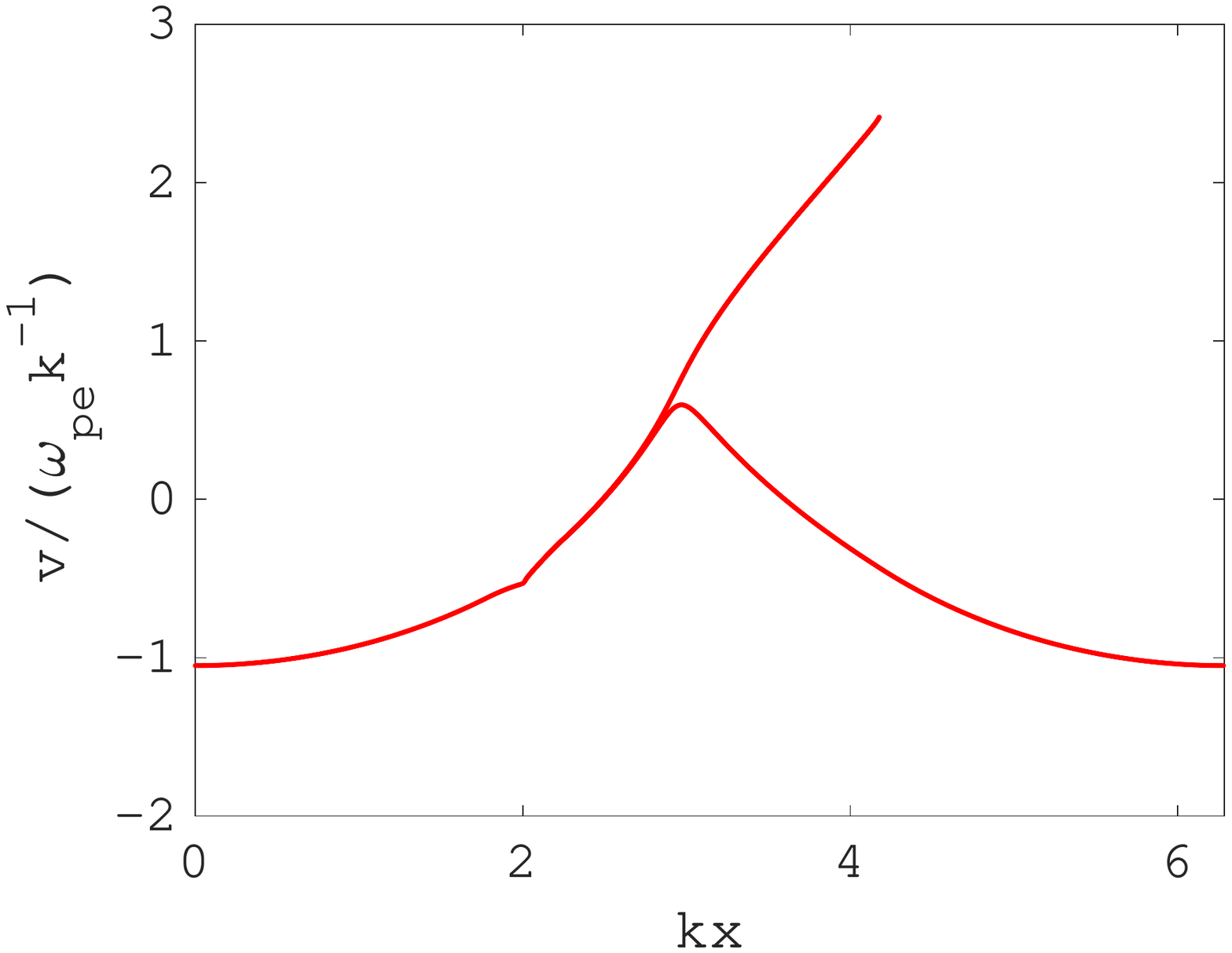}
\vspace{-1.4in}
\caption{Phase-space snap shot at $\omega_{pe}t = \pi/2$ for $A = 1.05$.}
\label{fig:figure1}
\end{minipage}
\hspace{0.5cm}
\begin{minipage}[b]{0.45\linewidth}
\centering
\includegraphics[width=\textwidth]{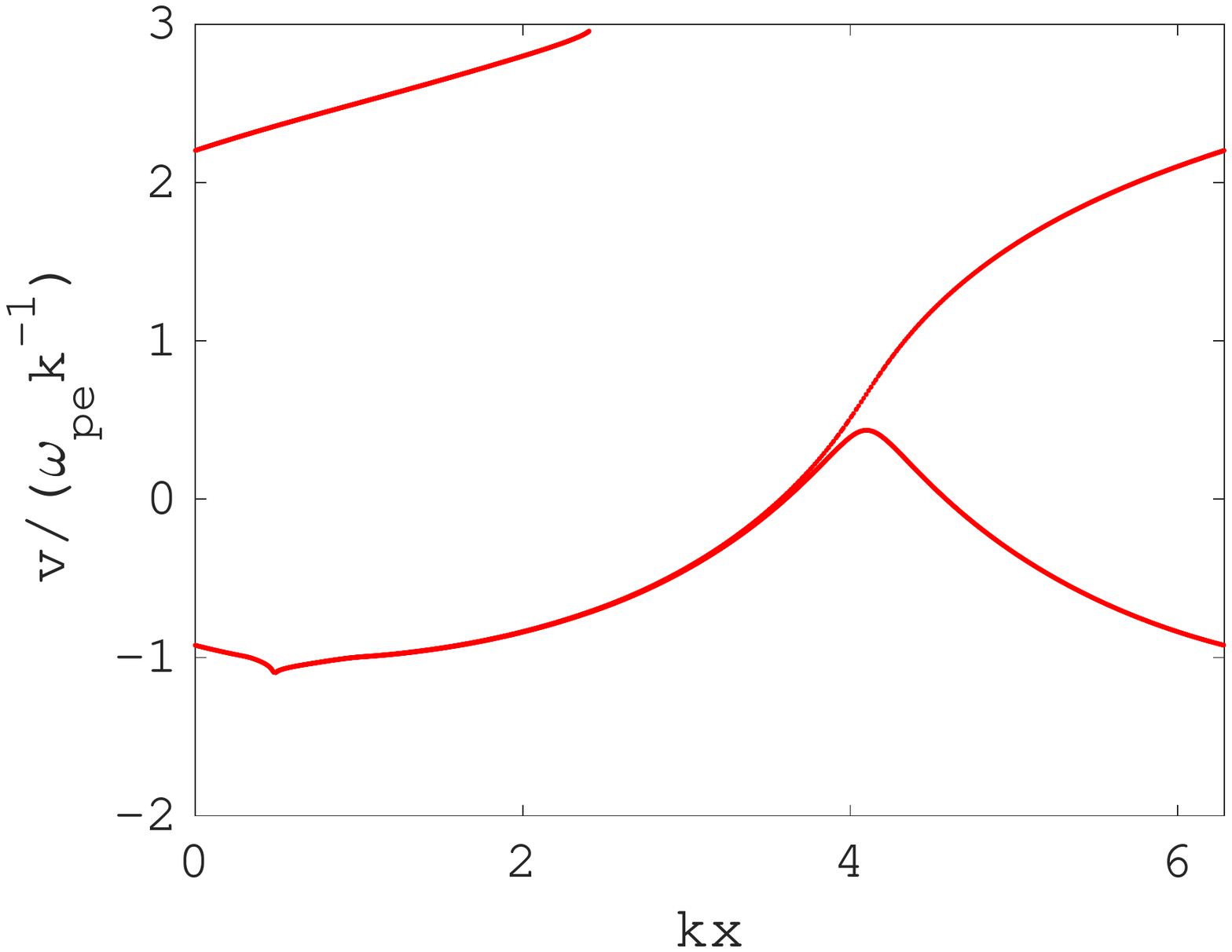}
\vspace{-1.4in}
\caption{Phase-space snap shot at $\omega_{pe}t = \pi$ for $A = 1.05$.}
\label{fig:figure2}
\end{minipage}
\end{figure}

\begin{figure}[h!]
\begin{minipage}[b]{0.45\linewidth}
\centering
\includegraphics[width=\textwidth]{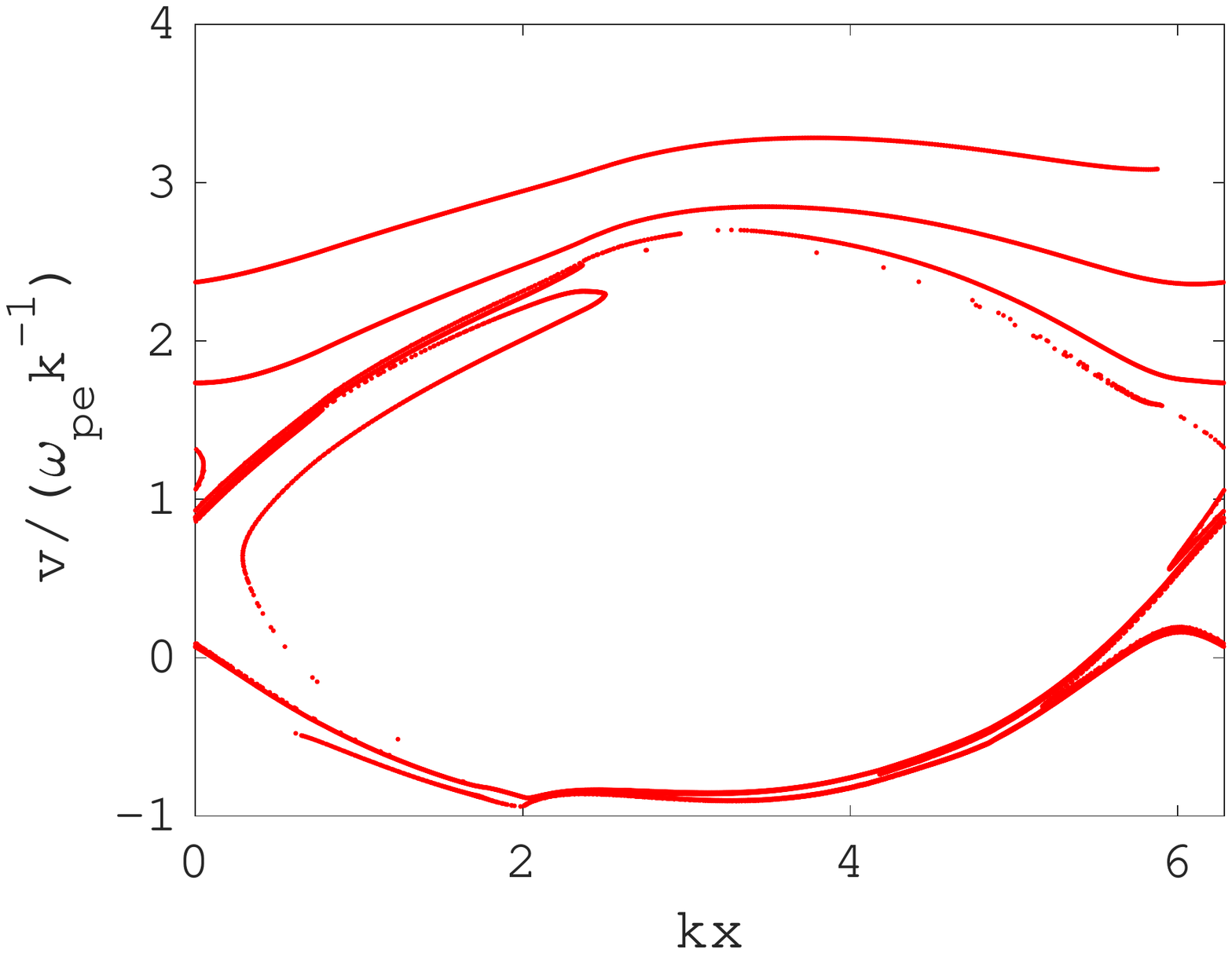}
\vspace{-1.4in}
\caption{Phase-space snap shot at $\omega_{pe}t = 9\pi/2$ for $A = 1.05$.}
\label{fig:figure9}
\end{minipage}
\hspace{0.5cm}
\begin{minipage}[b]{0.45\linewidth}
\centering
\includegraphics[width=\textwidth]{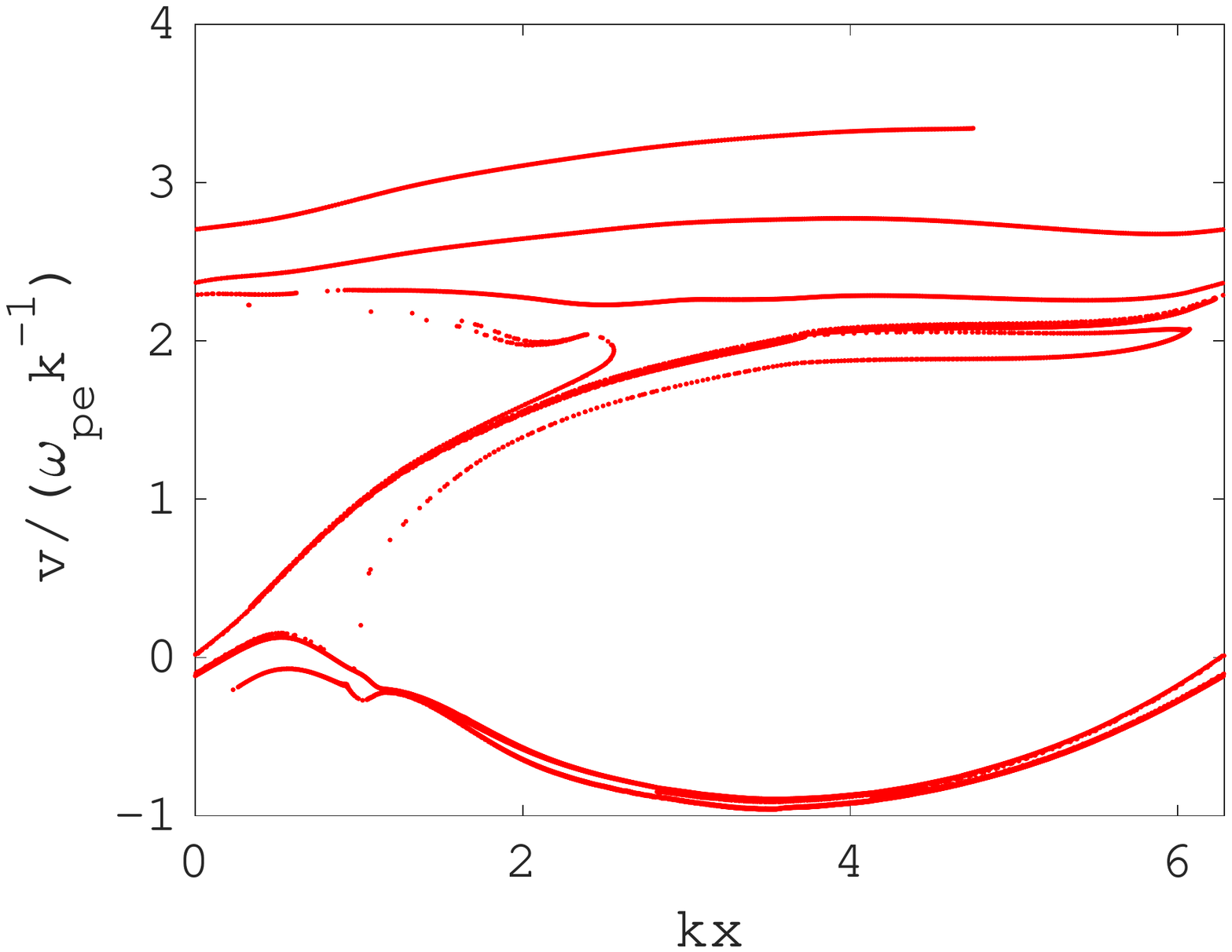}
\vspace{-1.4in}
\caption{Phase-space snap shot at $\omega_{pe}t = 5\pi$ for $A = 1.05$.}
\label{fig:figure10}
\end{minipage}
\end{figure}

\begin{figure}[h!]
\begin{minipage}[b]{0.45\linewidth}
\centering
\includegraphics[width=\textwidth]{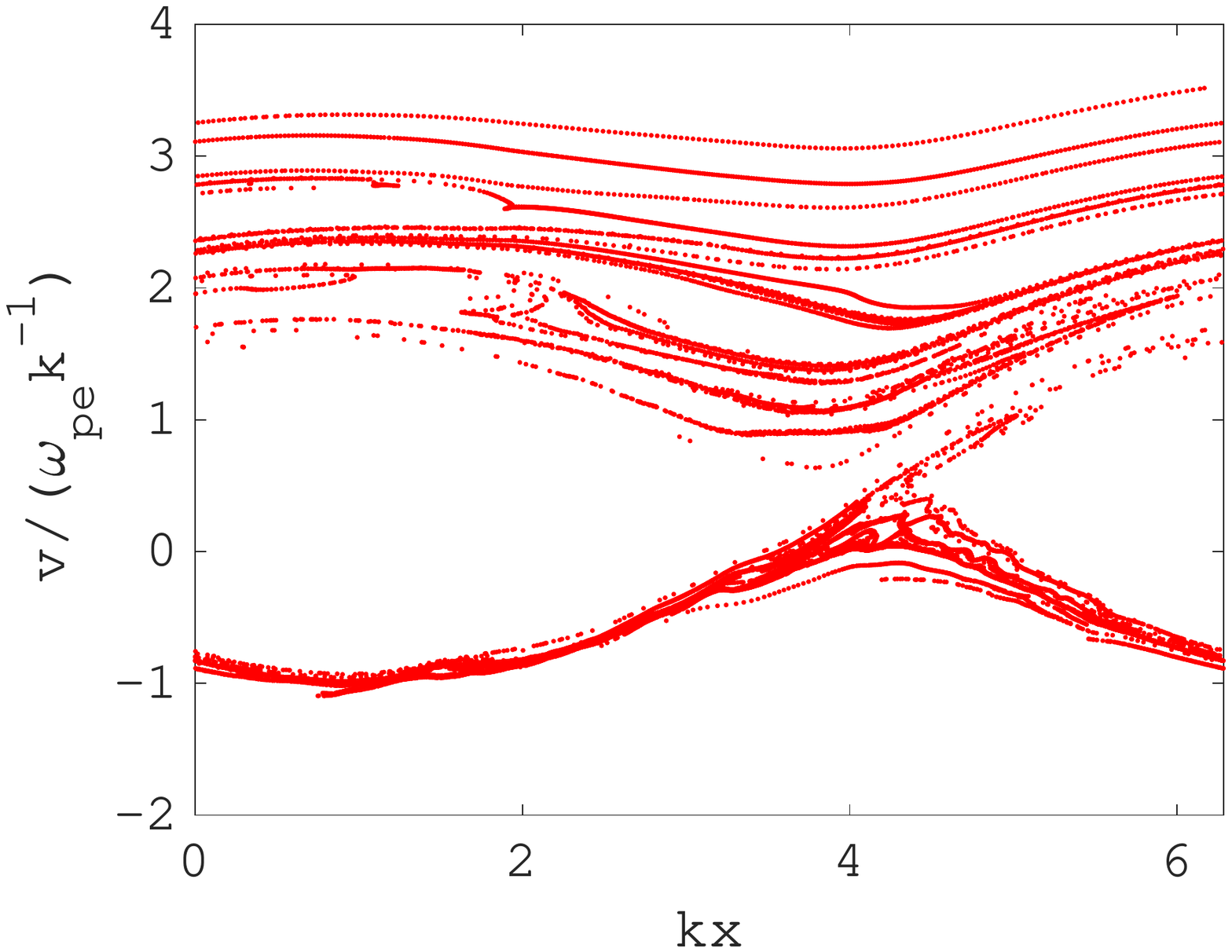}
\vspace{-1.4in}
\caption{Phase-space snap shot at $\omega_{pe}t = 21\pi/2$ for $A = 1.05$.}
\label{fig:figure11}
\end{minipage}
\hspace{0.5cm}
\begin{minipage}[b]{0.45\linewidth}
\centering
\includegraphics[width=\textwidth]{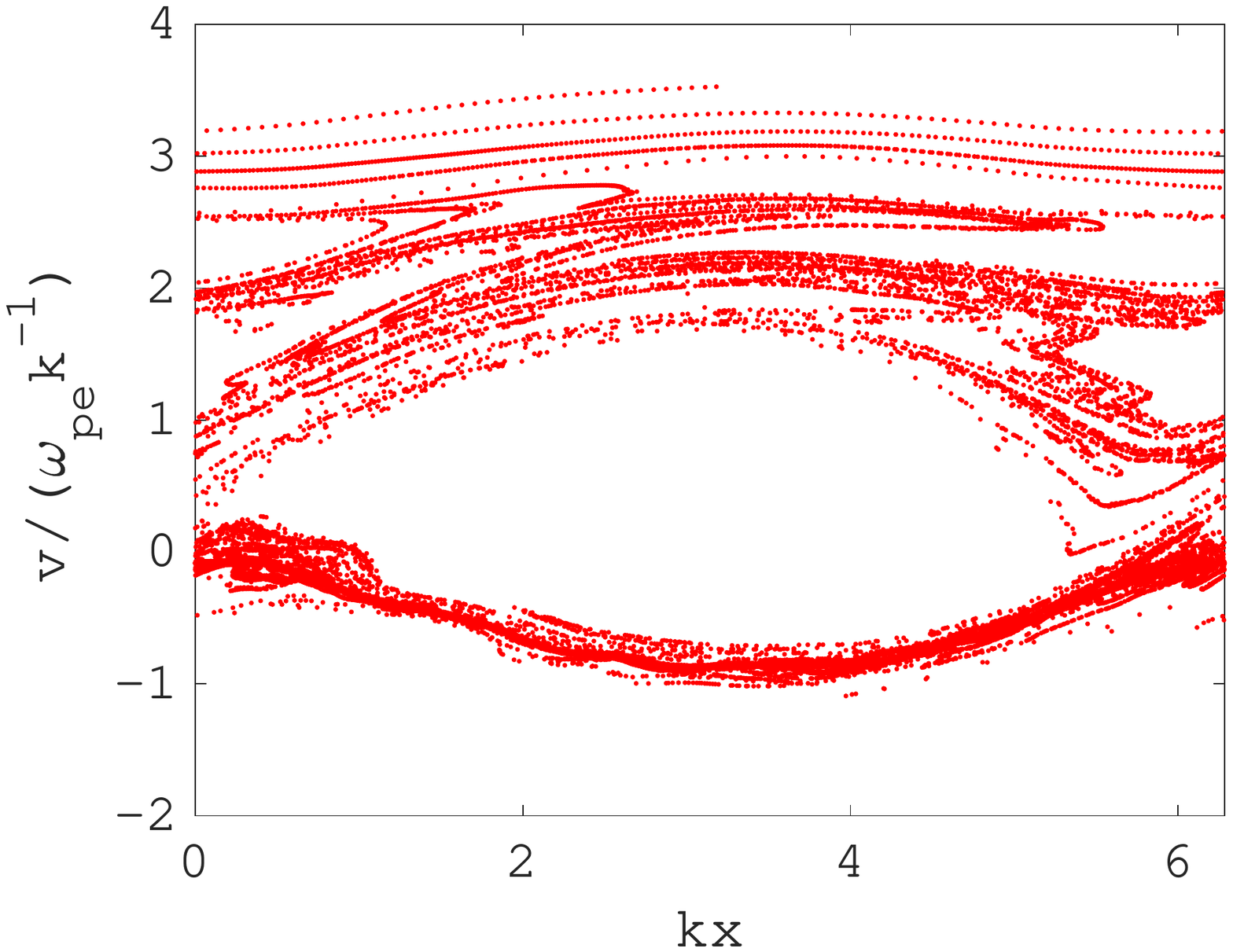}
\vspace{-1.4in}
\caption{Phase-space snap shot at $\omega_{pe}t = 31\pi/2$ for $A = 1.05$.}
\label{fig:figure12}
\end{minipage}
\end{figure}

\begin{figure}[h!]
\begin{minipage}[b]{0.45\linewidth}
\centering
\includegraphics[width=\textwidth]{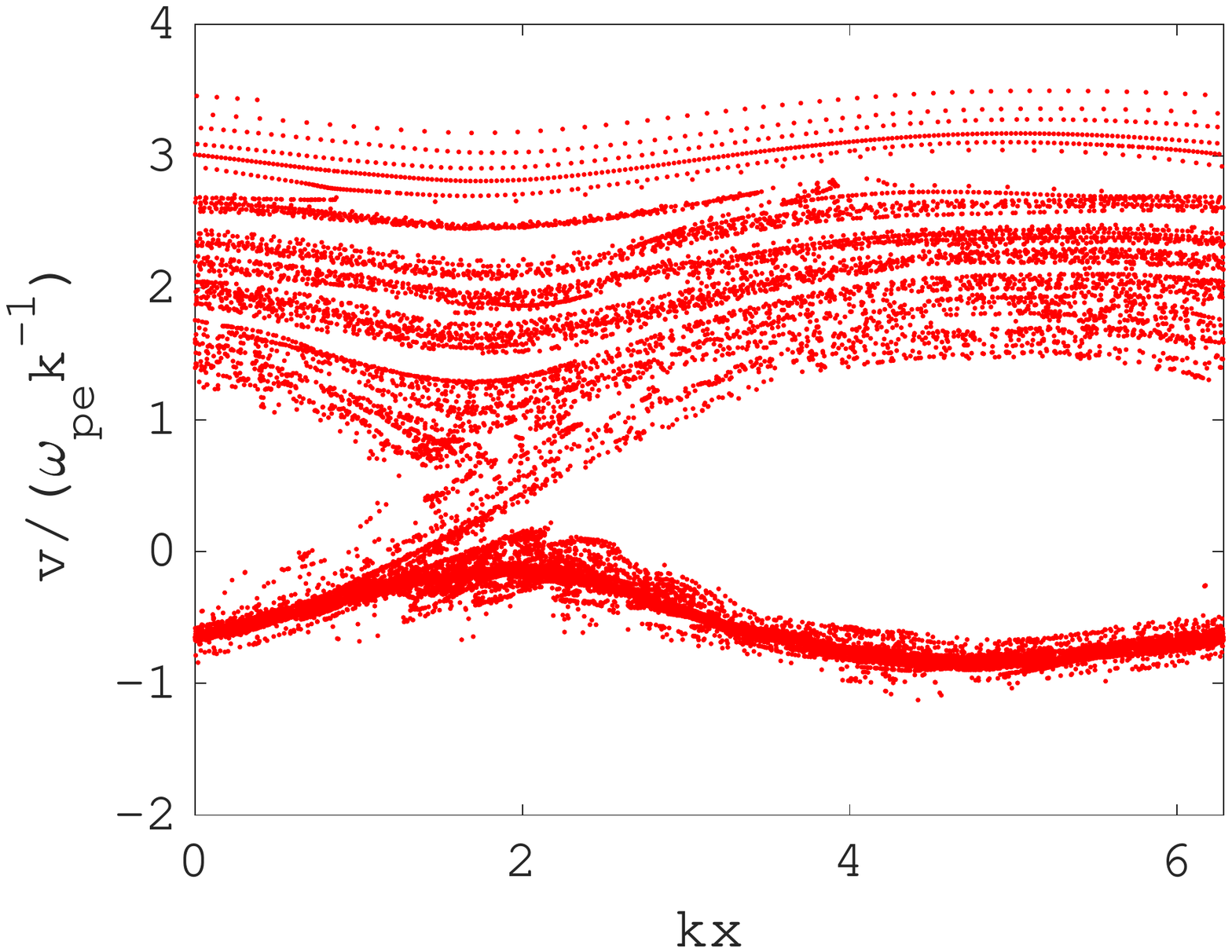}
\vspace{-1.4in}
\caption{Phase-space snap shot at $\omega_{pe}t = 41\pi/2$ for $A = 1.05$.}
\label{fig:figure13}
\end{minipage}
\hspace{0.5cm}
\begin{minipage}[b]{0.45\linewidth}
\centering
\includegraphics[width=\textwidth]{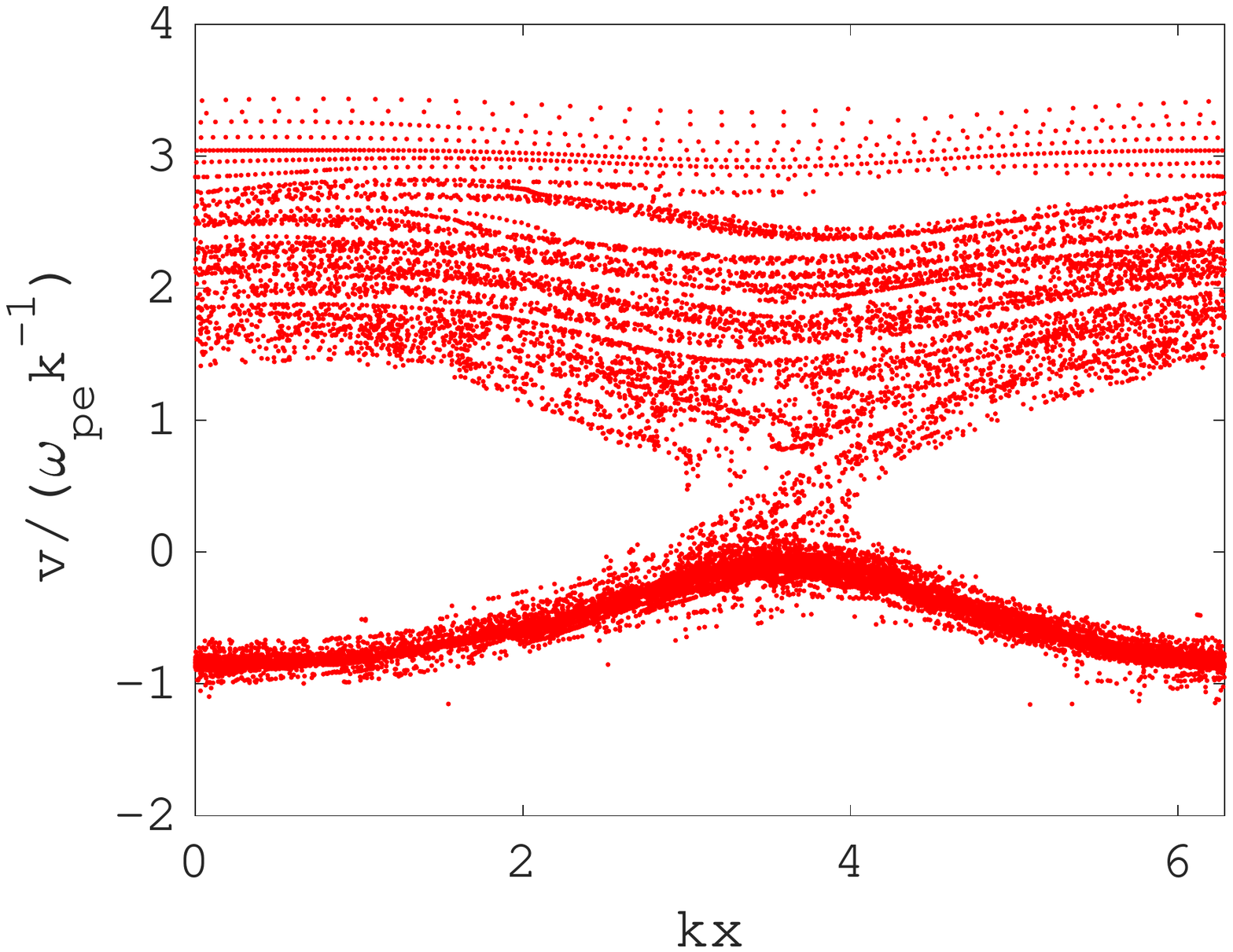}
\vspace{-1.4in}
\caption{Phase-space snap shot at $\omega_{pe}t = 51\pi/2$ for $A = 1.05$.}
\label{fig:figure14}
\end{minipage}
\end{figure}

\begin{figure}[h!]
\begin{minipage}[b]{0.45\linewidth}
\centering
\includegraphics[width=\textwidth]{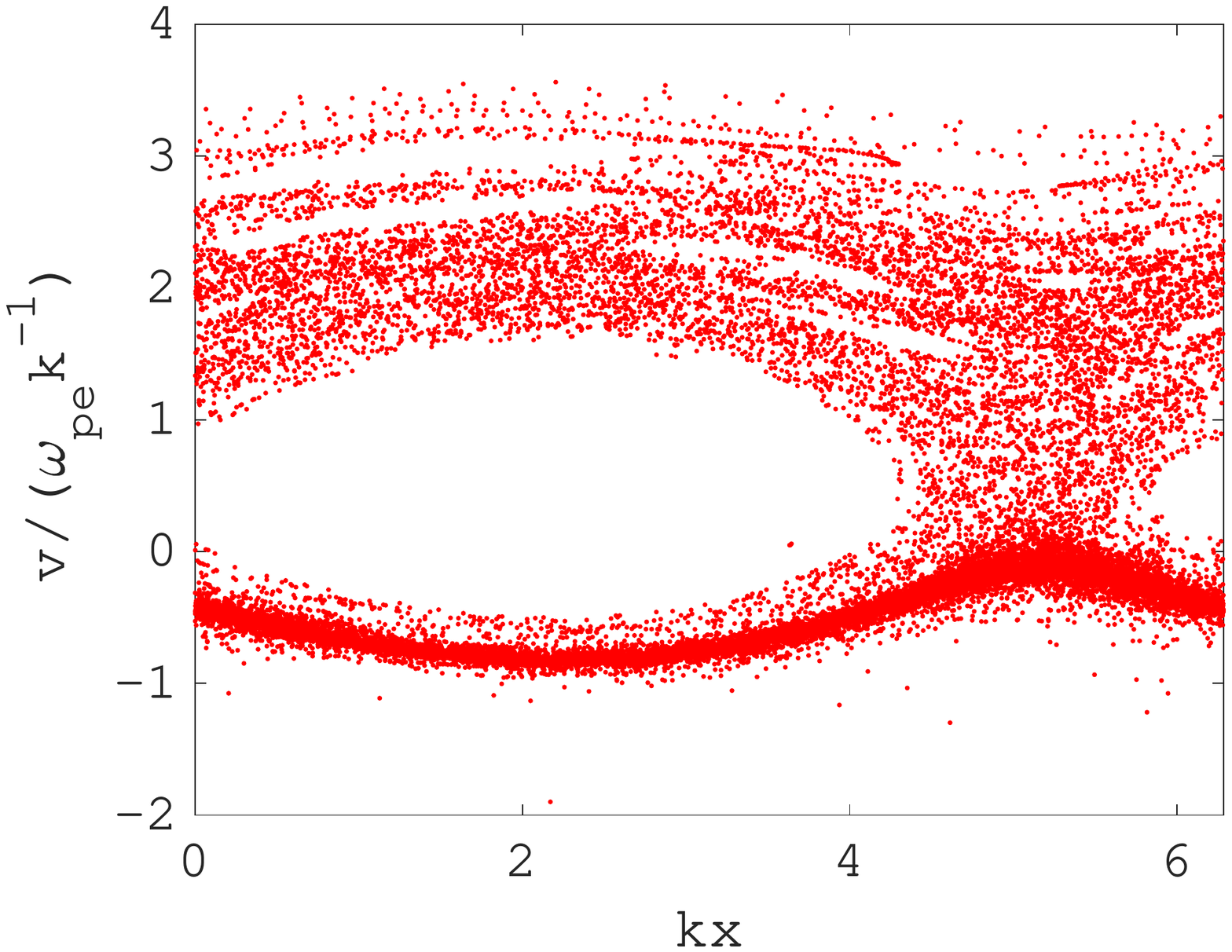}
\vspace{-1.4in}
\caption{Phase-space snap shot at $\omega_{pe}t = 101\pi/2$ for $A = 1.05$.}
\label{fig:figure15}
\end{minipage}
\hspace{0.5cm}
\begin{minipage}[b]{0.45\linewidth}
\centering
\includegraphics[width=\textwidth]{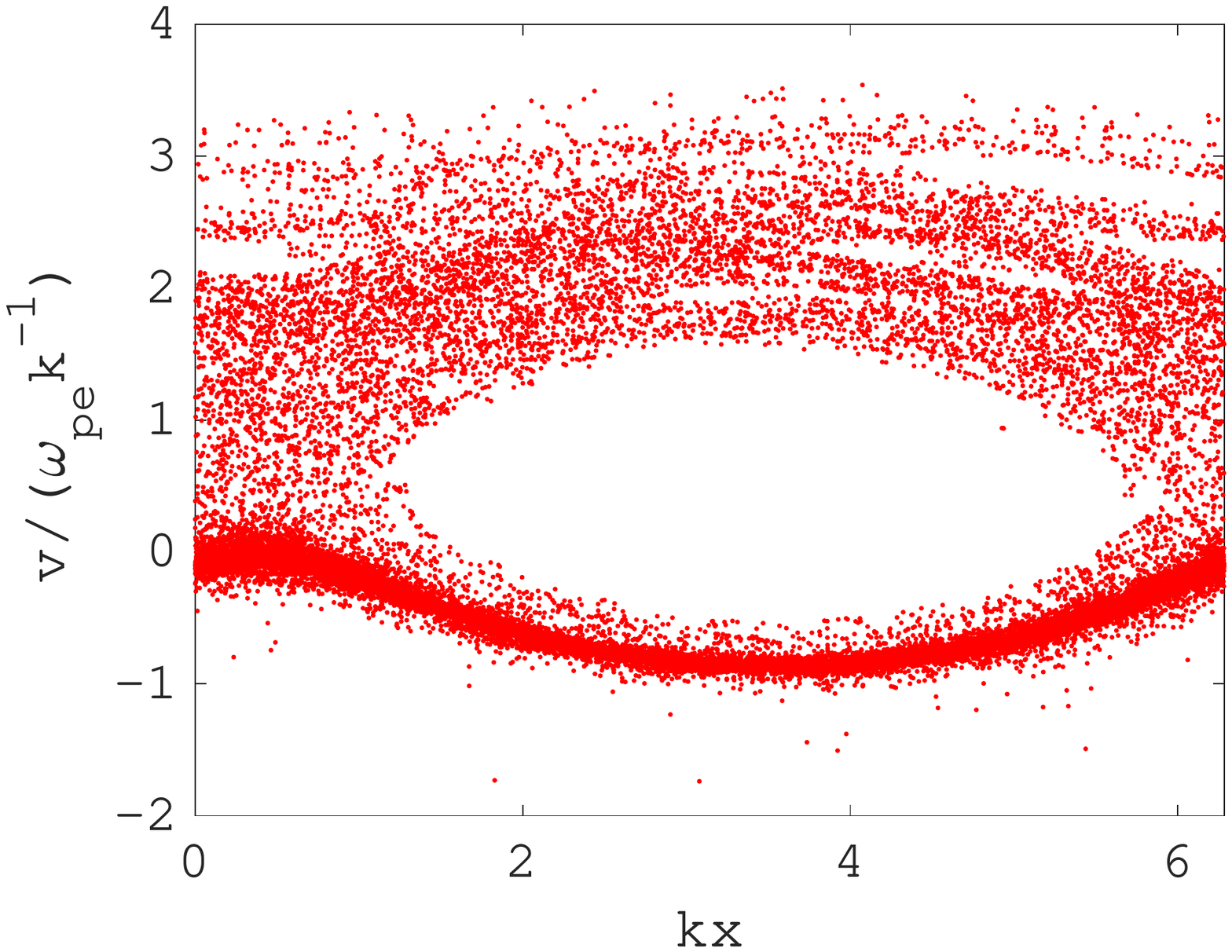}
\vspace{-1.4in}
\caption{Phase-space snap shot at $\omega_{pe}t = 151\pi/2$ for $A = 1.05$.}
\label{fig:figure16}
\end{minipage}
\end{figure}

Here we carry out 1-D PIC simulation \cite{birdsall2004plasma}  
with periodic boundary conditions, in order to study the evolution of
large amplitude cold plasma BGK modes \cite{davidson1968nonlinear,davidson1972methods,albritton1975relation} beyond the wave breaking amplitude. Our simulation parameters are as
follows: total number of particles (N) $\sim$ 4$\times10^4$, number of grid points (NG) $\sim$ 4$\times10^3$,
time step $\Delta t$ $\sim$ $\pi/50$. 
Ions are assumed to be infinitely massive which are just providing
the neutralizing background to the electrons. Normalization is as follows.
$ x \rightarrow kx$, $t \rightarrow \omega_{pe}t$, $n_e \rightarrow n_e/n_0$,
$v_e \rightarrow v_e/(\omega_{pe}k^{-1})$ and $E \rightarrow keE/(m\omega_{pe}^2)$, where
$\omega_{pe}$ is the plasma frequency and $k$ is the wave number of the longest (fundamental) mode.
System length is $L$ = 2$\pi$, therefore $k = 1$. The cold plasma BGK modes are initiated in the simulation 
by the following set of initial conditions,

\begin{equation}
 \xi = k A \cos kx_{eq}, v_e = \omega_{pe}k^{-1} A \sin kx_{eq},
\end{equation}
where $\xi$ is the initial displacement from the equilibrium position ($x_{eq}$) of the electrons, 
$v_e$ is the initial velocity and `$A$' is the amplitude of the perturbation. When `$A$' is kept below 
$1$, these initial conditions lead to propagating cold BGK-waves at the phase velocity $\omega_{pe}/k$ $\sim 1$. As an 
example, we choose a 
case when $A = 1.05$ which is slightly beyond the critical (breaking) amplitude. In figures~\ref{fig:figure1}-\ref{fig:figure16} we show 
snap-shots of phase space at various at time from $\omega_{pe}t = \pi/2$ to $\omega_{pe}t = 151\pi/2$.   

At $A = 1.05$, electron-plasma wave starts to see some particle traveling close to its phase velocity, 
as a result these particles take energy from 
the wave and get accelerated to higher velocities as shown in figures~\ref{fig:figure1}-\ref{fig:figure14}.   
After approximately $25$ plasma periods system reaches a stationary state because   
the phase space does not evolve any further as is shown in figures~\ref{fig:figure15}-\ref{fig:figure16}.
In figures~\ref{fig:figure17}-\ref{fig:figure18} we show the temporal evolution of averaged electrostatic 
energy (ESE) and averaged kinetic energy (KE) respectively. This evolution is found to be consistent with the 
phase space evolution as ESE initially decreases and later (approximately after 25 plasma periods) 
saturates at a finite amplitude. These results are found to be consistent with the previous study \cite{verma2012residual} 
as after the breaking of cold plasma BGK modes all the coherent energy does not vanish 
but some fraction always remains with the wave.   

However, when we look at the distribution function in figure~\ref{fig:figure19} at $\omega_{pe}t = 195\pi/2$, 
a flattening about $v = 0.5\omega_{pe}k^{-1}$ is noticed. This indicates that frequency of the remnant BGK-wave is 
much smaller than the electron plasma frequency $\omega_{pe}$ which is further confirmed from the 
fast Fourier transform (FFT) of the temporal evolution of the electric field at a fixed point in space, as 
is shown in figure~\ref{fig:figure20}. Moreover, it is also observed that frequency of the remnant 
BGK-wave decreases with increasing initial 
perturbation amplitude `$A$' as is shown in figure~\ref{fig:figure21}. This behaviour is contrary to the previous work 
\cite{verma2012residual} where    
frequencies of the remnant waves were never found to be below electron plasma frequency. 

Since, the distribution function acquires a finite width in the stationary state, this indicates that plasma 
gets heated due to wave-breaking. 
The effective temperature is measured numerically by calculating 
the second moment of the final distribution function, which
gives the root mean square speed $v_{rms}$ of the particles, and
define an effective thermal velocity $v_{th}$ of the particles, which
is found to be $\sim 1.0113 \omega_{pe}k^{-1}$. We have also measured the temperature of fast and slow electrons 
(electrons supporting the wave) separately and it is found that the temperature of the slow electrons 
remains more or less same (corresponding thermal velocity $\sim 0.35 \omega_{pe}k^{-1}$), however, temperature of the fast 
electrons increases with increasing initial 
amplitude `$A$' as we see in figure~\ref{fig:figure22}. 

In figure~\ref{fig:figure25} we show the final electric field as a function of initial amplitudes which indicates 
that amplitude of the saturated electric field decreases much faster at comparatively smaller initial amplitudes. 
Note here that we are unable to compare the amplitude of final electric field with Coffey's 
theoretical expression \cite{coffey1971breaking,mori1990wavebreaking} 
of maximum electric field in warm plasma because $v_{th} / v_{ph} > 1$ and this   
gives imaginary values when used in Coffey's expression. 

In the next section we provide a physical interpretation of the numerical experiments.      
\section{Interpretation of the results}
\begin{figure}[h!]
\begin{minipage}[b]{0.45\linewidth}
\centering
\includegraphics[width=\textwidth]{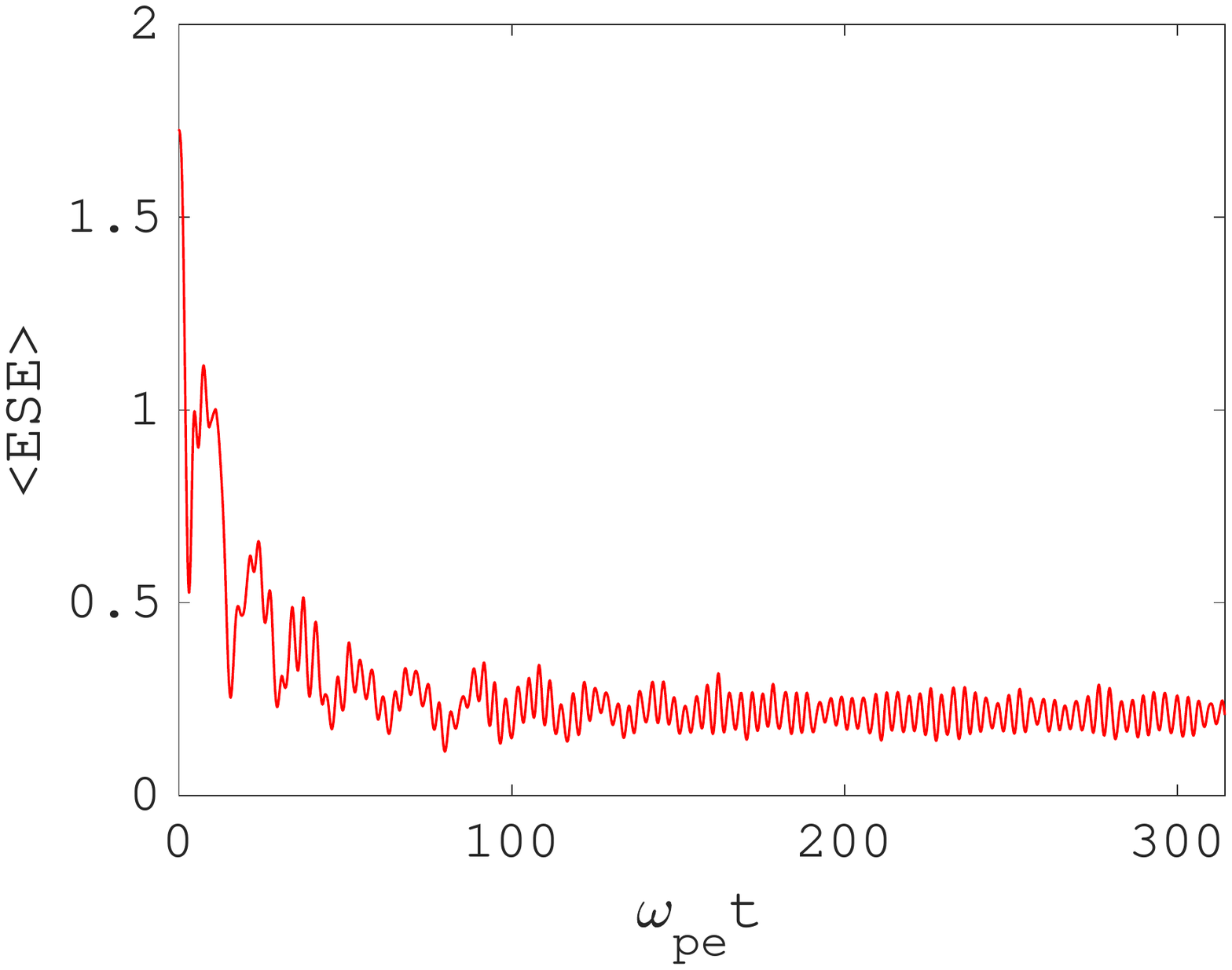}
\vspace{-1.4in}
\caption{Temporal evolution of averaged electrostatic energy $<ESE>$ for $A = 1.05$.}
\label{fig:figure17}
\end{minipage}
\hspace{0.5cm}
\begin{minipage}[b]{0.45\linewidth}
\centering
\includegraphics[width=\textwidth]{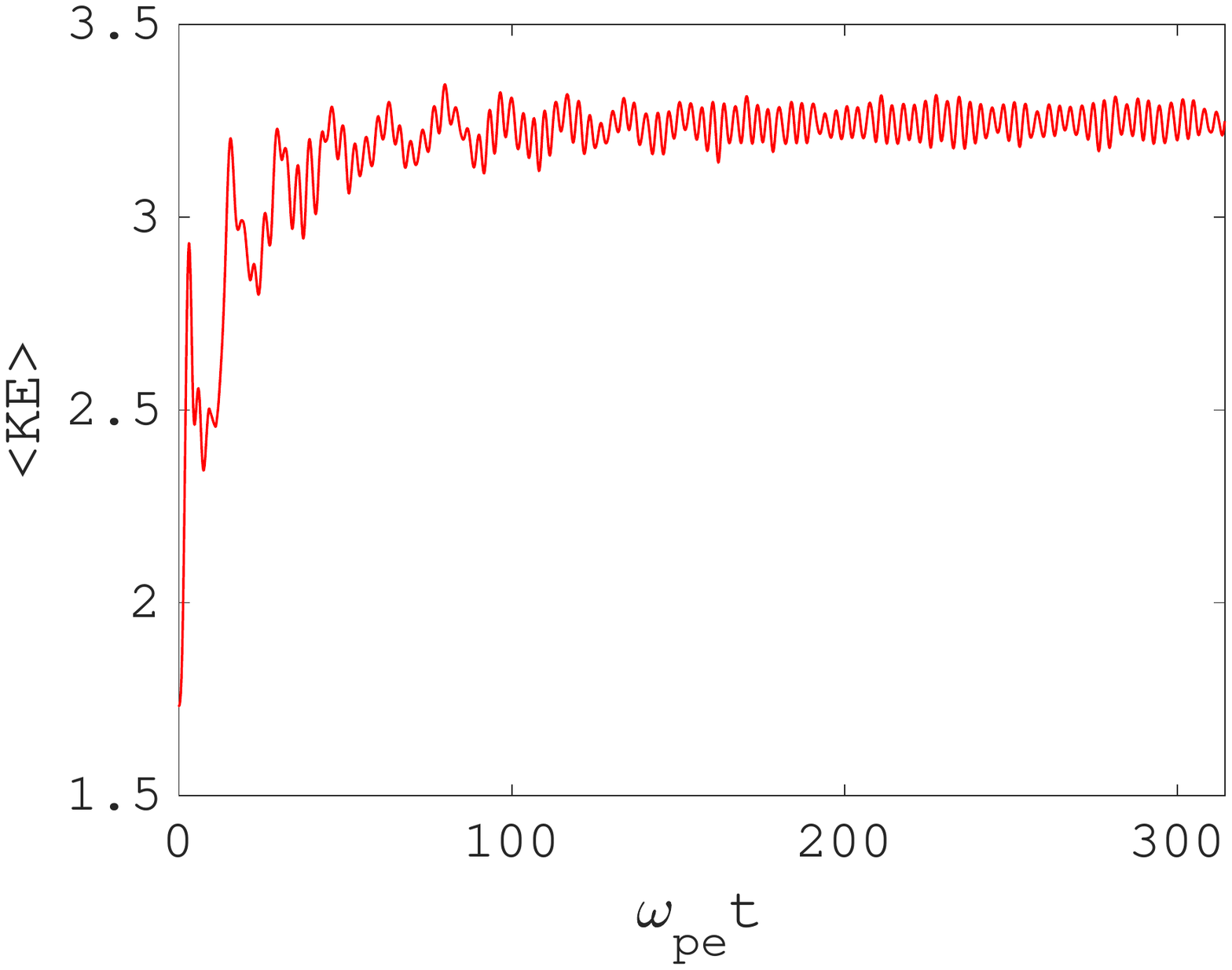}
\vspace{-1.4in}
\caption{Temporal evolution of averaged kinetic energy $<KE>$ for $A = 1.05$.}
\label{fig:figure18}
\end{minipage}
\end{figure}

\begin{figure}[h!]
\includegraphics[height=6in,width=5in]{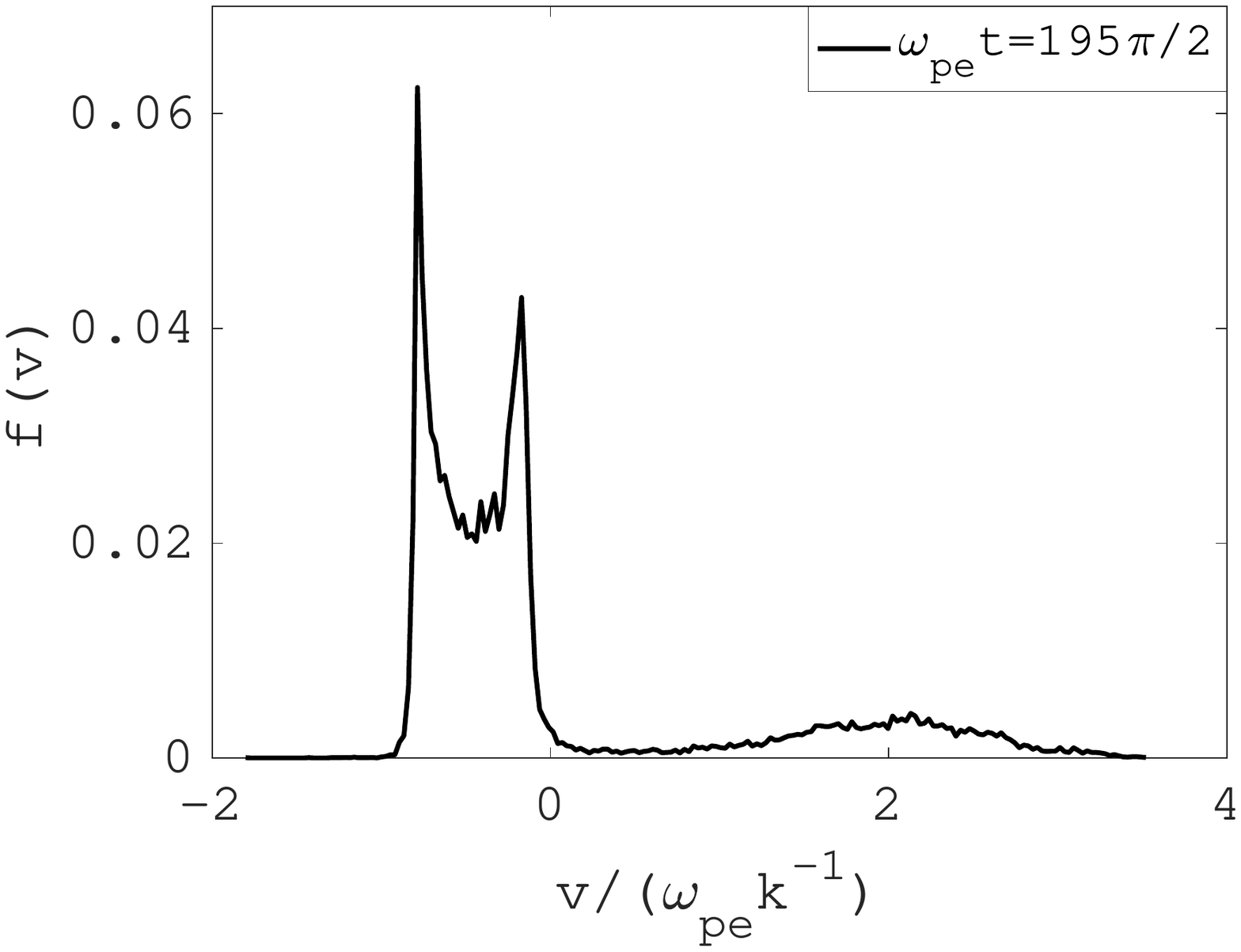}
\vspace{-1.5in}
\caption{Distribution function at $\omega_{pe}t = 195\pi/2$ for $A = 1.05$.}
\label{fig:figure19}
\end{figure}

\begin{figure}[h!]
\includegraphics[height=6in,width=5in]{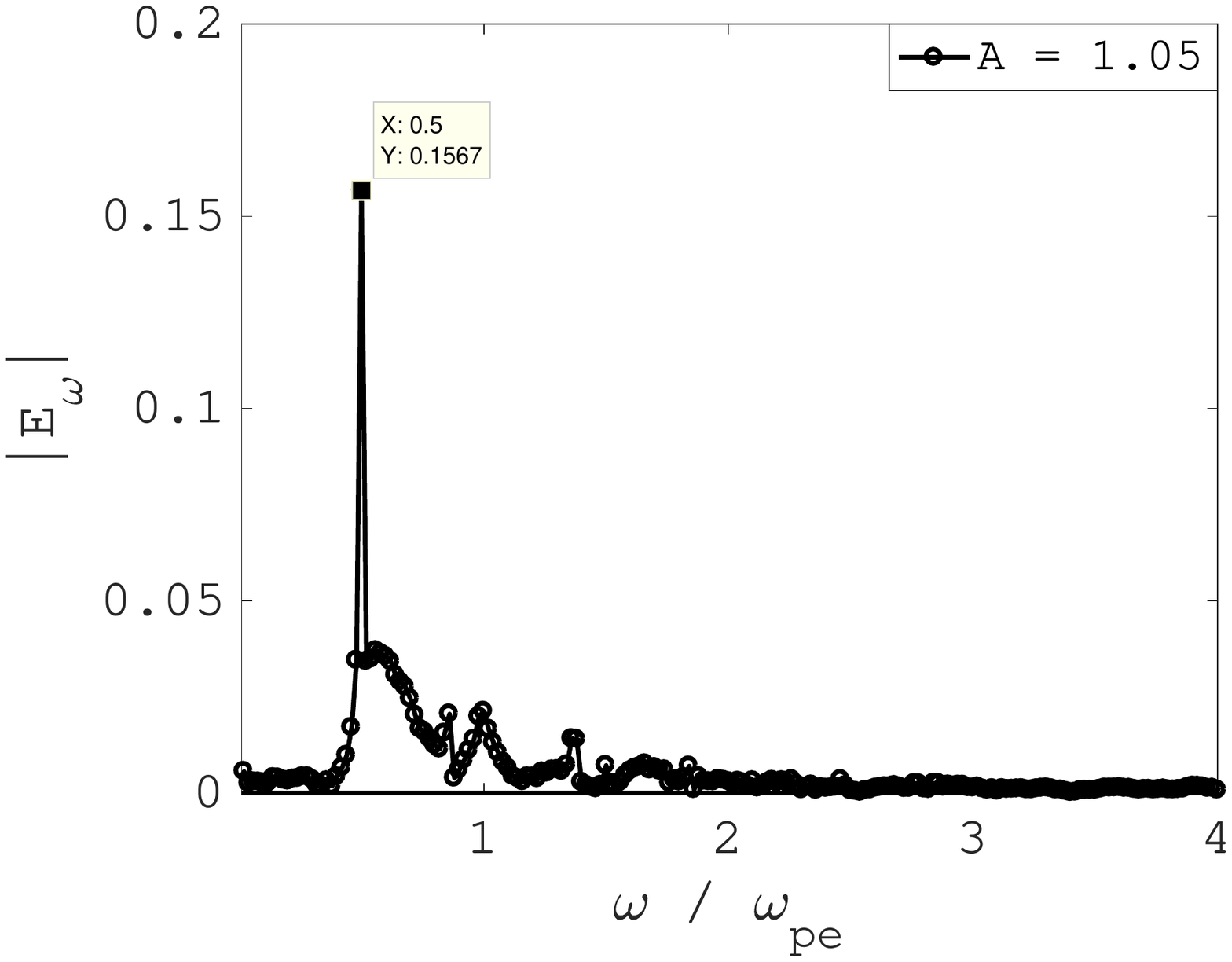}
\vspace{-1.5in}
\caption{FFT of the temporal profile of the electric field for $A = 1.05$.}
\label{fig:figure20}
\end{figure}

\begin{figure}[h!]
\includegraphics[height=6in,width=5in]{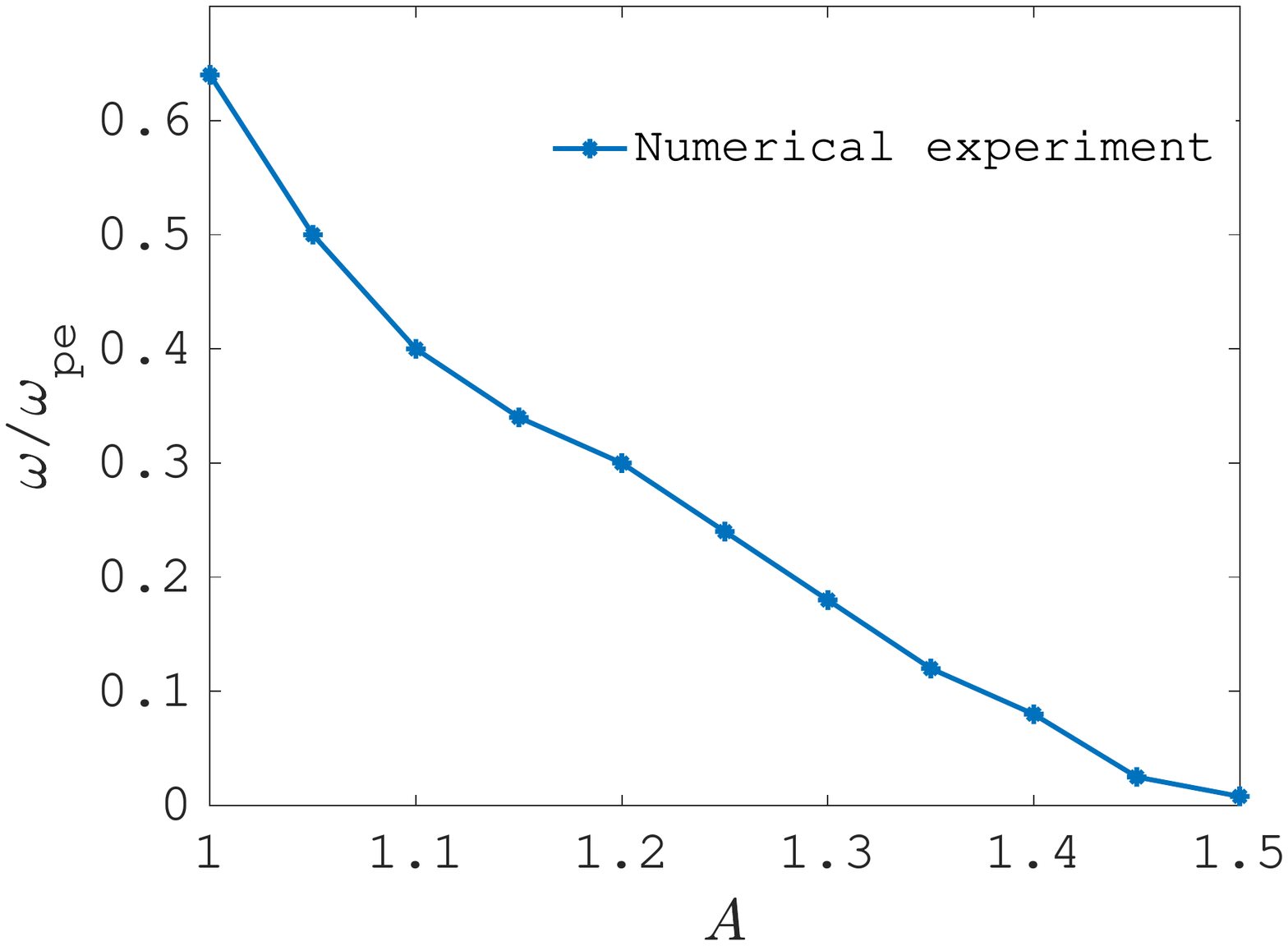}
\vspace{-1.5in}
\caption{Frequency of the remnant BGK wave versus initial amplitude at $\omega_{pe}t = 195\pi/2$.}
\label{fig:figure21}
\end{figure}

\begin{figure}[h!]
\includegraphics[height=6in,width=5in]{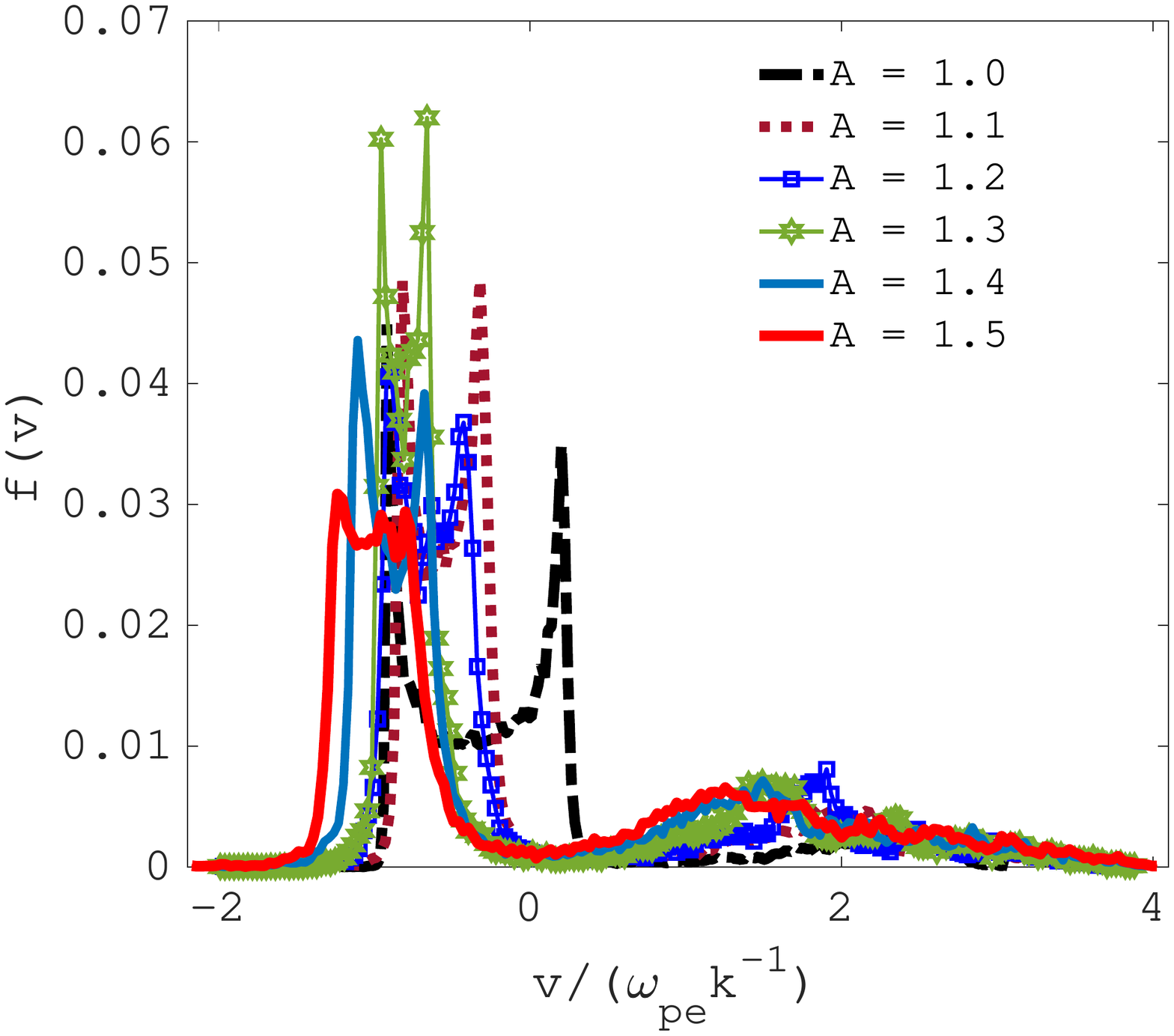}
\vspace{-1.5in}
\caption{Distribution functions for different initial amplitudes at $\omega_{pe}t = 195\pi/2$.}
\label{fig:figure22}
\end{figure}

\begin{figure}[h!]
\includegraphics[height=6in,width=5in]{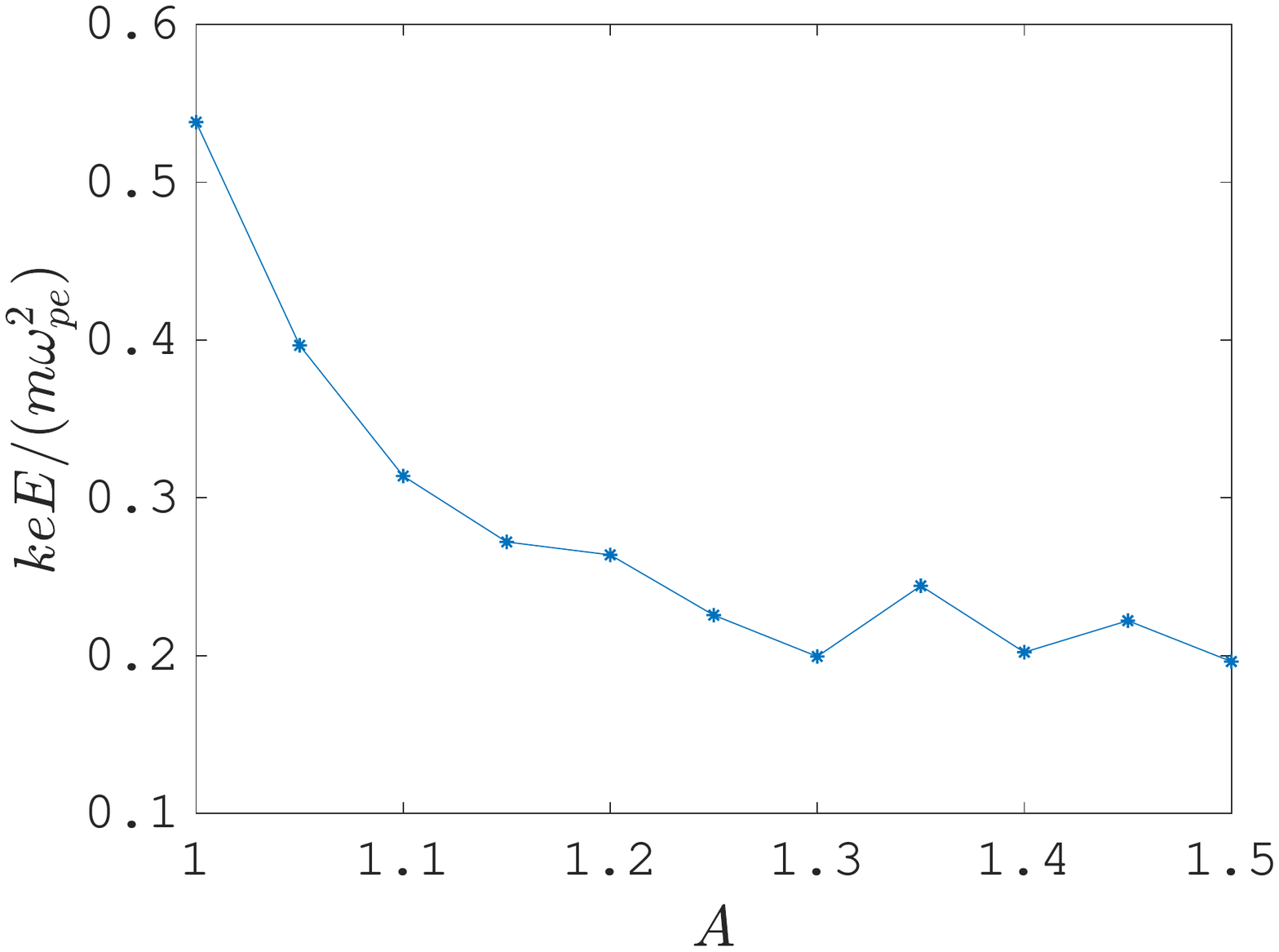}
\vspace{-1.5in}
\caption{Amplitude of the saturated electric field as a function of initial amplitude at $\omega_{pe}t = 195\pi/2$.}
\label{fig:figure25}
\end{figure}

At an amplitude $A = 1.05$ even at $\omega_{pe} t = 0$ the cold BGK wave is able to see some electrons with velocities 
close to its phase velocity ($\sim 1$).  Therefore, it traps and accelerates  
them to high energies within one plasma period as we see in figures~\ref{fig:figure1}-\ref{fig:figure2}. 
We call these high energetic electrons as fast electrons because they do not oscillate at the electron-plasma frequency 
and their velocities are higher than the phase velocity of the cold plasma wave ($\sim \omega_{pe}/k$). 
These fast electrons  
give rise to a mean drift velocity along the $+x$-direction. Note here that there is no 
net acceleration along the opposite direction which can cancel out the latter mean drift velocity.  
Therefore, in order to conserve the total momentum, rest of the plasma which is oscillating coherently 
starts to acquire a mean drift along the $-x$ direction. However, from the linear dispersion relation 
it is well known that if we have a plasma wave propagating in the $+x$-direction along with a 
mean drift in the $-x$-direction (say, $-v_0$), phase velocity (and hence frequency $\omega$) 
of the plasma wave reduces due to the Doppler effect, {\it i.e.} $\omega = \omega_{pe} - k v_0$. 
Thus an increase in the mean drift velocity of the fast electrons introduces a negative shift in the  
frequency and hence in the phase velocity of the electron-plasma wave. 
This enables plasma wave to trap 
and accelerate comparatively slower (than $\sim \omega_{pe}/k$) particles to higher velocities and thereby further 
reducing the phase velocity, as is seen in figures~\ref{fig:figure9}-\ref{fig:figure10}. 
This process keeps going on until the amplitude of the accelerating field, which is decreasing as 
the electrostatic energy is being lost in accelerating electrons, becomes so small that it is no more 
able to increase the mean drift velocity of the fast electrons any further.  
Therefore, no more further reduction of the phase velocity.                      
Although, in this process  
few electrons with wave frame kinetic energy less than the potential energy 
maxima get trapped in the potential well of the plasma wave, which can then exchange
energy with the wave during trapped oscillations.  
Meanwhile, with the progress of time, the trapped particle distribution becomes
a well phase mixed through nonlinear Landau damping \cite{o1965collisionless} 
such that an asymptotic state is finally reached where the
distribution function becomes stationary in its own frame
and the ESE neither grows nor damps. This explains the saturation in the ESE and KE in figures~\ref{fig:figure17}-\ref{fig:figure18}
after a certain time. These states are known as BGK waves in warm plasma which are time-stationary solutions of Vlasov-Poisson system  
\cite{bernstein1957exact,manfredi1997long,hutchinson2017electron} and the  
amplitude of such wave depends on the plateau $\Delta v_{trap}$, the width over which electrons are
trapped in the wave trough which, in the normalized unit, is given by $\Delta v_{trap} = 2 \sqrt{(2 \phi)}$. In this case, $\Delta v_{trap}$ is approximately 
$1.68$ which is measured from figures~\ref{fig:figure17}-\ref{fig:figure18} and 
amplitude of the wave potential $\phi$ is $0.357$ which together satisfy the above mentioned theoretical expression quite 
accurately.      

Note here that in the previous work \cite{verma2012residual} we have studied the breaking of standing 
plasma waves where no decrease in the frequency of the remnant waves was observed because 
there were fast electrons drifting in both $\pm x$-directions thereby annihilating the net drifting 
effect on the frequency of the plasma oscillations.  

In the next section we construct a simplified theoretical model in order to gain more insight into the numerical 
results. 
\section{Theoretical model} 
\begin{figure}[h!]
\includegraphics[height=6in,width=5in]{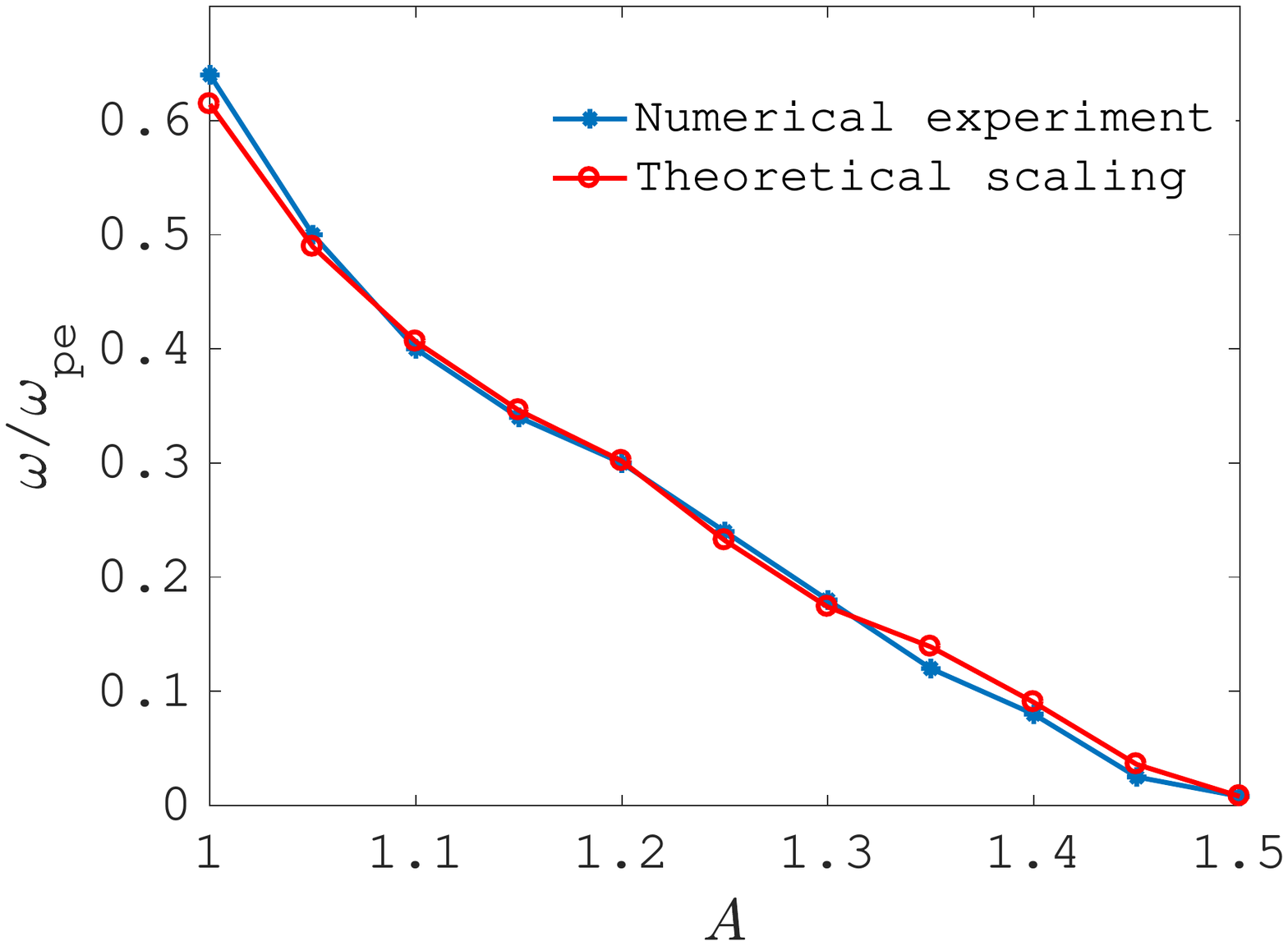}
\vspace{-1.5in}
\caption{Frequency of the remnant BGK wave ($\omega$) as function of initial amplitude `$A$' -- a comparison between numerical 
experiment (blue line points) and theoretical scaling (red line points). }
\label{fig:figure26}
\end{figure}

In the stationary state we have two kind of electrons. Ones which are oscillating coherently, however the second ones are 
fast electrons drifting along the $+x$-direction. 
These fast electrons are responsible for inducing a mean drift velocity along the $-x$-direction in the rest of the plasma 
such that total mean momentum remains zero so as to satisfy momentum conservation. One can write down 
a linear dispersion relation for such a system as follows, 
\begin{eqnarray}\label{disp1}
1 - \frac{\omega_{p1}^2}{(\omega-kv_{01})^2} - \frac{\omega_{p2}^2}{(\omega-kv_{02})^2} = 0,
\end{eqnarray}
where $\omega_{p1}$ and $\omega_{p2}$ are the frequencies of the coherently oscillating electrons and  
fast electrons respectively and $v_{01}$, $v_{02}$ are their respective mean drift velocities. 
Thermal corrections have been ignored here for simplicity. 
Note here that $v_{02} > \omega / k $ and $\omega_{p2}^2 < \omega_{p1}^2$ as the number of fast electrons being much 
smaller than the number of coherently oscillating electrons. Therefore, the third term can be dropped 
from equation~\eqref{disp1} because it has a 
weaker contribution as compared to the second term. Equation~\eqref{disp1} is therefore approximated as,
 \begin{eqnarray}\label{disp2}
1 - \frac{\omega_{p}^2}{(\omega-kv_{01})^2} \approx 0,\hspace{0.1cm} i.e. \hspace{0.5cm} 
\omega \approx \omega_{p} + kv_{01}
\end{eqnarray}
where $\omega_{p}$ $\approx \omega_{p2}$ is the frequency of the cold plasma wave. Since $v_{01}$ arises in order 
to balance the effect of $v_{02}$, it is always 
negative and increases for higher initial amplitudes. This explains why the remnant wave acquires a frequency below 
$\omega_{pe}$ and why it decreases with increasing initial amplitude. In figure~\ref{fig:figure26} we provide a comparison 
between numerical results and scaling for the theoretical model which shows a good agreement between the two. Here $v_{01}$ 
is measured as a mean velocity of the coherently oscillating electrons.
 \section{Summary and discussion}
In this paper, we have studied the long time evolution of large amplitude 
non-relativistic cold plasma BGK waves in the 
wave breaking regime and found that the plasma gets warm after the wave-breaking, however a fraction of 
initial energy always remains with the wave in the form of BGK-modes in warm plasmas \cite{bernstein1957exact,hutchinson2017electron}. 
These findings are in agreement 
with the previous work where breaking of large amplitude standing plasma oscillations was studied \cite{verma2012residual}. 
However, in contrary  
the frequency of the remnant BGK-wave has been found to be below electron-plasma frequency which decreases 
further with increasing initial amplitude. This is because the fast electrons which are generated after the wave-breaking 
acquire a mean drift in the direction of acceleration and this mean drift has been found to increase for larger initial amplitudes. 
In order to cancel out this effect so as to conserve the total momentum, bulk of the plasma 
(coherently oscillating electrons) acquires a negative mean drift velocity. 
The latter drift velocity allows remnant wave to acquire a frequency well below electron plasma frequency.
However, in the previous study the electrons were drifting in both $\pm x$-direction thereby canceling out 
the drifting effect on the bulk of the plasma. Therefore, we have not observed any decrease in the frequency of 
the coherent plasma oscillations after the wave-breaking \cite{verma2012residual}.   
 
Furthermore, the acceleration process after the wave-breaking has also been 
found to be different from the previous study \cite{verma2012residual}. This is due to the fact that in the case of standing 
oscillations we do not have any particles with velocities close to the phase velocity $\sim \omega_{pe}/k$. But, soon after 
the wave-breaking some high `$k$' modes (with phase velocities much smaller than $\sim \omega_{pe}/k$) 
acquire a small amount of energy which they resonantly transfer to the comparatively 
much slower particles and accelerate them to comparatively higher velocities. With increased velocities these particles 
get an opportunity to resonantly 
interact with smaller `$k$' modes (which have comparatively higher energies) and get accelerated to further higher velocities (for more detailed description please 
refer \cite{verma2012residual}). However, in the present study 
we already have electrons with velocities close to $\sim \omega_{pe}/k$ from the beginning, so they get accelerated to 
much higher velocities with in 
a plasma period. And these accelerated electrons self-consistently introduce a negative shift in the frequency of 
the plasma wave so that it can resonantly interact and give energy to comparatively slower particles. 
This process keeps going on until a stationary state is reached.        

In order to gain an insight into the physics of the observed results 
in the numerical experiments, we have provided a theoretical model based on linear dispersion relation which has shown 
a good agreement between the two.    
Note here that although the frequencies of these BGK-waves are below electron-plasma frequency $\omega_{pe}$, they are not electron 
acoustic modes \cite{sircombe2006aspects, chakrabarti2009nonlinear} because these modes are being generated 
due to the Doppler effect not due to the shielding effect \cite{chakrabarti2009nonlinear}.   
These modes seem to 
be similar to KEEN waves as reported in \cite{johnston2009persistent}.        

These investigations have direct relevance in particle acceleration experiments because we learn here that  
plasma waves get slower after the wave-breaking in order to conserve the momentum and this may 
also be true for relativistic wake wave-breaking \cite{modena1995electron}.


\pagebreak 


\end{document}